\documentclass[
  prb,
  twocolumn,
  showpacs,
  amsmath,
  amssymb,
  superscriptaddress, 
]{revtex4}

\usepackage{bm}
\usepackage{graphicx}

\begin{document}

\title{Equilibration in a chiral Luttinger liquid}

\author{I.\ V.\ Protopopov}
\affiliation{
 Institut f\"ur Nanotechnologie, Karlsruhe Institute of Technology,
 76021 Karlsruhe, Germany
}

\affiliation{
 L.\ D.\ Landau Institute for Theoretical Physics RAS,
 119334 Moscow, Russia
}

\author{D.\ B.\ Gutman}
\affiliation{Department of Physics, Bar Ilan University, Ramat Gan 52900,
Israel }

\author{A.\ D.\ Mirlin}

\affiliation{
 Institut f\"ur Nanotechnologie, Karlsruhe Institute of Technology,
 76021 Karlsruhe, Germany
}
\affiliation{
 Institut f\"ur Theorie der Kondensierten Materie,
 Karlsruhe Institute of Technology, 76128 Karlsruhe, Germany
}

\affiliation{
 Petersburg Nuclear Physics Institute,  188300 St.~Petersburg, Russia.
}

\begin{abstract}
We explore the weak-strong-coupling Bose-Fermi duality in a model of a single-channel integer or fractional quantum Hall edge state with a finite-range interaction. The system is described by a chiral Luttinger liquid with non-linear dispersion of bosonic and fermonic excitations. We use the bosonization, a unitary transformation, and a refermionization to map the system onto that of weakly interacting fermions at low temperature $T$ or weakly interacting bosons at high $T$. We calculate the equilibration rate which is found to scale with temperature as $T^5$ and $T^{14}$ in the high-temperature (''bosonic'') and the low-temperature (''fermonic'') regimes, respectively. The relaxation rate of a hot particle with the momentum $k$  in the fermonic regime scales as $k^7T^7$.  

\end{abstract}

\pacs{
}

\maketitle

\section{Introduction}
Quantum kinetics of one-dimensional (1D) interacting systems is a hot topic in the contemporary condensed matter physics. 
It is now actively explored both theoretically and experimentally. On experimental side, both fermionic (carbon nanotubes, metallic and semiconductor nanowires, quantum Hall edges) and 
bosonic (cold atoms in optical traps) systems are studied. The continuing progress in the sample fabrication technology allows the study of the quantum kinetics and relaxation in a variety of setups; see, in particular, Refs.~\onlinecite{Mason09,Yacoby10,Altimiras10,Paradiso11,Deviatov11,Prokudina2014,Kinoshita06,Hofferberth07}.

On the theoretical side,  a common starting point for the description  of 1D interacting  systems is the Tomonaga-Luttinger model of fermions with linear dispersion relation interacting via point-like interaction. By virtue of bosonization\cite{Stone_book, Delft,Gogolin, Giamarchi} the Tomonaga-Luttinger model can be mapped  to the free bosons with 
linear spectrum which is in turn equivalent to free fermions via refermionization\cite{Matis_Lieb1965, Rozhkov,imambekov09, imambekov11}. 

Being a linear theory, Tomonaga-Luttinger model does not allow for the relaxation to thermal equilibrium. It is by now well understood that the energy relaxation in clean 1D systems occurs due  to terms 
that are  irrelevant in the renormalisation group (RG) sense, which makes  the calculation of the corresponding relaxation rates rather complicated.   
In the recent years various perturbative schemes were developed to address the kinetics in interacting 1D systems. Generally, they can be divided into two classes - ''fermonic'' and ''bosonic''. The  fermonic approaches start from the formulation of the theory either in terms of original fermions\cite{Khodas2007, Lunde07, micklitz11, ristivojevic13, Dmitriev2012} (assumed to be weakly interacting) or in terms of appropriately defined fermionic quasiparticles\cite{Protopopov14, MatveevFurusaki2013, Gangardt2014}. 
One can then speak about the lifetime of fermionic excitations in the system and discuss the kinetics in the framework of the fermionic kinetic equation. The approaches of the second class appeal to the bosonic description of 1D system, discuss the lifetimes of bosonic excitations and employ the bosonic kinetic equation as a tool to address kinetics. This idea has been used for the analysis of strongly-interacting fermions (with Luttinger liquid parameter $K\ll 1$) forming a state which is close to Wigner crystal\cite{Lin2013, apostolov13}.    

It was pointed out recently that fermionic and bosonic languages are dual\cite{Protopopov2014_2}.
The two typical perturbations to the Tomonaga-Luttinger model are the finite curvatures of fermionic and bosonic 
spectra. In the model of 1D fermions interacting via finite rage interaction they are controlled\cite{Protopopov2014_2} by the fermionic mass $m$ (correction to the linear fermionic dispersion $\delta\epsilon_k=k^2/2m$) and the radius of interaction $l$ (correction to the linear bosonic spectrum $\delta\omega_q=-u_0 l^2q^3$, where $u_0$ is the sound velocity), respectively. We note that the quadratic correction to the fermionic spectrum translates into cubic interaction of bosons in the bosonized version of the theory\cite{schick68}.
Thus, the interaction in the fermonic picture corresponds to the dispersion in the bosonic one and vice versa.  

The comparison of the non-linear corrections to the fermionic and bosonic dispersion relations reveals  an important parameter of the theory $\lambda_E=ml^2 E$ with $E$ being the typical energy of a single-particle excitation. This energy scale is set by temperature $T$ for a system near thermal equilibrium.  The stability of bosons with respect to their interaction (caused by fermionic dispersion) is controlled by the bending of the bosonic energy-momentum relation. 
Accordingly, for $\lambda_T\gg 1$ one expects that the bosonic form of the theory is the proper starting point for the perturbative treatment. In particular, in this ``bosonic regime'' one can speak of the lifetime of the bosonic excitations and employ the bosonic kinetic equation to describe the kinetics.  Conversely, small value of the parameter $\lambda_T$ means that (at the relevant  energy scale) we can view the system as consisting of weakly-interacting fermionic quasiparticles and the the fermionic kinetic equation can be used to analyze the kinetics. Remarkably, the same parameter $\lambda_E$ emerges from the analysis of the fermionic and bosonic descriptions of the dissipationless kinetics in a problem of propagation of a density pulse on top of zero-temperature background\cite{Protopopov13,Protopopov14}; in that case the energy scale $E$ is determined by the pulse amplitude.

The parameter $\lambda_E$ grows with energy. Thus, the picture drawn above  implies (as shown within a systematic analysis in Ref.~\cite{Protopopov2014_2}) that at comparatively high temperatures the kinetics and, in particular, the rate at which a perturbed system relaxes to thermal equilibrium, is governed by the bosonic kinetic equation and the lifetimes of the bosonic excitations, while low-temperature equilibration is controlled by the fermionic relaxation rates.  

Quantum Hall edge states represent a particularly important realization of interacting 1D system. The state-of-the-art engineering of quantum Hall devices allows to create structures with a high degree of control on systems parameters, including non-equilibrium conditions. Recent experiments examine the non-equilibrium Mach-Zehnder interferometry \cite{heiblum,roche,strunk,schoenenberger} and energy relaxation in quantum Hall edges \cite{Altimiras10,Deviatov11, Paradiso11,Prokudina2014}.

In this work we explore the Bose-Fermi duality outlined above in a single-channel quantum Hall edge state (integer or fractional) with a finite-range interaction, thus extending the results of  Ref.~\onlinecite{Protopopov2014_2} to the case of a chiral Luttinger liquid. We present detailed description of the relaxation processes in this system. Our analysis covers both the ``bosonic'' high-temperatures regime and the ``fermionic'' low-temperature one. We show that in these two regimes the characteristic equilibration rate  scales with temperature as $T^5$ and $T^{14}$, respectively. Such a dramatic change in the temperatures behavior should be amenable to experimental test. 

The paper is organized as follows. In Sec. \ref{sec:sec1} we fix notations and present our model of a quantum Hall edge. 
In Sec. \ref{sec:sec2} we consider the high-temperature regime and analyze the equilibration starting from the the bosonic description of the problem. Section \ref{sec:sec3} is devoted to the fermionic formulation of the theory and the analysis of equilibration at low temperatures.  Finally, Sec.~\ref{sec4} contains a summary and a discussion of our results.

\section{Effective theory of a single-channel quantum Hall edge}
\label{sec:sec1}

The effective low energy description of  a quantum Hall edge in terms of chiral bosonic fields was introduced by Wen \cite{Wen}. 
In the simplest situation of a single-channel edge, which is realized in $\nu=1$ integer quantum Hall state as well as in the $\nu=1/m$  Laughlin states, the theory involves 
a single chiral density field $\rho(x)$ with the commutation relation (in momentum domain)\cite{Remark:CommutatioRelation} 
\begin{equation}
 \left[\rho_q,\rho_{-q}\right]= \frac{L q}{2\pi}\ ,
 \label{Eq:sec1:commutation}
\end{equation}
and the Hamiltonian of the system is given by
\begin{equation}
 H_0=\frac{\pi u_0 }{ L}\sum_q :\rho_q \rho_{-q}:_B.
 \label{Eq:sec1:H0}
\end{equation}
Here $L$ is the total length of the edge, $u_0$ is the velocity of edge excitations and $::_B$ denotes the normal ordering with respect to bosonic creation and annihilation operators. 

The the bosonic Hamiltonian (\ref{Eq:sec1:H0}) can also be viewed as a Hamiltonian of free chiral fermions\cite{Remark:BoseFermiMapping}
\begin{equation}
 H_0=u_0\sum_k k :a^+_k a_k:_F,
\end{equation}
where  $::_F$ denotes the fermonic normal ordering. The mapping to the fermionic theory goes via the representation of the density 
operators $\rho_q$ as bilinear functions of fermions $a_k$
\begin{equation}
 \rho_q=\sum_k a^+_{k}a_{k+q}.
 \label{Eq:sec1:Refermionization}
\end{equation}

Hamiltonian  (\ref{Eq:sec1:H0}) was extremely fruitful for the analysis of various low-energy properties of QH edges. However, being an effective low-energy theory, Hamiltonian 
(\ref{Eq:sec1:H0}) is not exact. In a more general theory one expects corrections to Eq. (\ref{Eq:sec1:H0}) made out of various operators of higher scaling dimension. 
Phenomenologically one can express such a perturbed Hamiltonian as
\begin{equation}
 H=\frac{\pi u_0}{L}\sum_{q}\Gamma^{(2)}_q:\rho_q\rho_{-q}:_B+\pi u_0\sum_{n=3}H^{(n)}.
 \label{Eq:sec1:H}
 \end{equation}
The first term in Eq. (\ref{Eq:sec1:H}) describes free {\it dispersive} chiral boson with momentum-dependent velocity $u_q\equiv u_0\Gamma^{(2)}_q$ while the other terms represents  various $n$-boson interactions
\begin{equation}
 H^{(n)}=\frac{1}{L^{n-1}}\left(\frac{2\pi}{mu_0}\right)^{n-2}\sum_{\bf q}\Gamma^{(n)}_{\bf q}:\prod_{i=1}^n\rho_{q_i}:_B.
 \label{Eq:sec1:Hn}
\end{equation}
Here, ${\bf q}=(q_1, q_2,\ldots q_n)$, $\Gamma^{(n)}_q$ are dimensionless functions of momenta and $m$ is a phenomenological parameter with dimension of mass. 

In this paper we focus on the case of a finite-range interaction, thus assuming that the functions $\Gamma^{(n)}_{\bf q}$ are analytic at small momenta. 
The Taylor expansion of the function $\Gamma^{(2)}_q$
introduce into the problem a length-scale $l$ controlling the dispersion of our bosons
\begin{equation}
 \Gamma^{(2)}_q=1-q^2 l^2+\sum_{k=2}\gamma_{(2, 2k)}l^{2k}q^{2k}.
\label{Eq:sec1:Gamma2}
 \end{equation}
Here $\gamma_{(2, m)}$ are numerical coefficients. 

The higher-order vertices $\Gamma^{(n)}_{q}$  have Taylor expansions similar to Eq. (\ref{Eq:sec1:Gamma2}) and we assume them to be controlled by the same length scale $l$,
\begin{equation}
 \Gamma^{(n)}_{\bf q}=\gamma_{(n, 0)}+\sum_{k=1}\sum_{p_{n, 2k}}\gamma_{(n, p_{n, 2k})}l^{2k}p_{n, 2k}({\bf q}).
 \label{Eq:sec1:Gamman}
\end{equation}
The second sum in Eq. (\ref{Eq:sec1:Gamman}) runs over all linearly independent uniform $2k$-power symmetric polynomials of $n$ variables $(q_1, \ldots q_n)$.  
Subsequently we will need explicit expansions for the first few functions $\Gamma^{(n)}_{\bf q}$ and we present them here for future references:\cite{Remark:Gamma3}
\begin{eqnarray}
\Gamma^{(2)}_q &=& 1-l^2 q^2+\gamma_{(2, 4)}l^4 q^4+\gamma_{(2, 6)}l^6q^6+\ldots\ ;
\label{Eq:sec1:uq}
\\
 \Gamma^{(3)}_{\bf q} &=& \gamma_{(3, 0)}+\gamma_{(3, 2)}l^2\sum_{i=1}^{3}q_i^2+\gamma_{(3, 4)}l^4\left(\sum_{i=1}^{3}q_i^2\right)^2+\ldots\ ; \nonumber \\ && \\
 \Gamma^{(4)}_{\bf q} &=& \gamma_{(4, 0)}+\gamma_{(4, 2)}l^2\sum_{i=1}^{4}q_i^2+\ldots\ ;\\
 \Gamma^{(5)}_{\bf q} &=& \gamma_{(5, 0)}+\ldots\ .
 \label{Eq:sec1:gamma5}
\end{eqnarray}

The physical meaning of the phenomenological parameter $m$  introduced in Eq. (\ref{Eq:sec1:Hn}) becomes transparent if we note that the leading 
(in the RG sense) perturbation in the Hamiltonian (\ref{Eq:sec1:H}) is given by 
\begin{equation}
\delta H=\frac{2\pi^2\gamma_{(3, 0)}}{m}\int dx :\rho^3(x):_B. 
\end{equation}
Under the refermionization mapping (\ref{Eq:sec1:Refermionization}) this correction transforms into
\begin{equation}
 \delta H=\frac{3\gamma_{(3, 0)}}{2m}\sum_k k^2 :a_k^+a_k:_F.
 \label{Eq:sec1:FermionicCurvature}
\end{equation}
The parameter $m$ and the corresponding length scale ($1/m u_0$) describe thus the bending of the ``electronic'' spectrum in the system.

Equations (\ref{Eq:sec1:H}),(\ref{Eq:sec1:Hn}), (\ref{Eq:sec1:Gamma2}), and (\ref{Eq:sec1:Gamman}) fully describe  our phenomenological Hamiltonian of a quantum Hall edge valid at low momenta,
$q\ll 1/l$ and $q \ll m u_0$, and constitute the starting point of our subsequent analysis.  

Despite the fact that the perturbations introduced in the Hamiltonian $H$ are irrelevant in the RG sense, they have an important impact on the the kinetic properties of the edge.  In particular, the competition between  the fermionic curvature term (\ref{Eq:sec1:FermionicCurvature}) and the  bending of the  bosonic spectrum, 
accounted in the leading order by
\begin{equation}
 -\frac{\pi u_0 l^2}{L}\sum_{q}q^2:\rho_q\rho_{-q}:_B
\end{equation}
drives  the Fermi-Bose crossover\cite{Protopopov13, Protopopov14} in the dissipationless dynamics of a density perturbation. The later analysis was limited   to the time scales that are shorter than the relaxation rates studied in the present paper. 

Besides deciding on the fermionic or bosonic nature of the problem the  perturbative  terms in the Hamiltonian (\ref{Eq:sec1:H}) lead to the  finite lifetimes of  the bosonic and fermionic excitations. 
This aspect  of the problem is in the focus of the present work.
 
We conclude this section,  by stating the assumptions used below.
First, the expansion of the Hamiltonian in powers of density and momentum [Eqs. (\ref{Eq:sec1:H}) and (\ref{Eq:sec1:Gamman})] are meaningful provided that the characteristic energy scale in the problem (set by the temperature $T$) satisfies
\begin{equation}
 T\ll m u_0^2\,, \qquad Tl/u_0\ll 1.
 \label{Eq:sec1:TCond}
\end{equation}
Second, the parameter $l$ (that can be interpreted as the radius of interaction in the fermionic version of the theory)  is  assumed  to be large compared to the effective Fermi wavelength in the problem, $l\gtrsim 1/m u_0$. Note that this condition also allows the parameter $\lambda_T=ml^2 T$ to vary from large to small values within the temperature range set by 
Eq (\ref{Eq:sec1:TCond}).

\section{Bosonic lifetime}
\label{sec:sec2}

In this section we analyze the decay time of the bosonic excitations in the chiral edge described by the Hamiltonian (\ref{Eq:sec1:H}). More precisely, we consider the system at finite temperature $T$ and a "test"  boson with momentum $q_1\gtrsim  T/u_0$. We are interested in the decay rate of this bosonic excitation which is given by the scattering-out part of the corresponding scattering integral. 

\begin{figure}
\includegraphics[width=220pt]{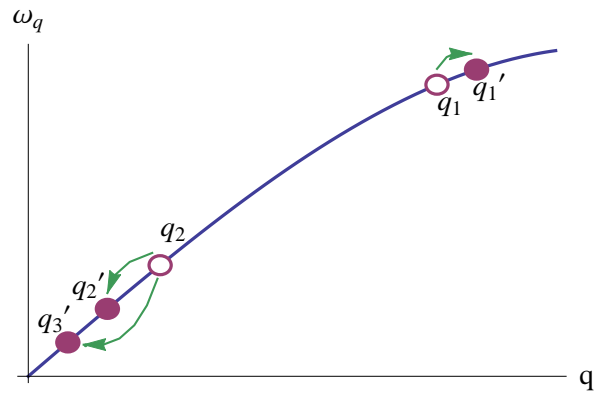}
\caption{\small The dominant bosonic scattering process contributing to the relaxation of a hot boson with momentum $q_1\gg T/u_0$.  The change in momentum of the hot boson, 
$q_1-q_1^\prime \propto T^3/u_0^3q_1^2$, is small compared to the typical momenta of the thermal bosons 
$q_2$, $q_2^\prime$ and $q_3^\prime$.  
}
\label{Fig:BosonicScattering}
\end{figure}

The simplest  process contributing to the decay of the state under consideration is shown in Fig. \ref{Fig:BosonicScattering}.
 It involves the scattering of the test boson 
by the thermal boson with  momentum $q_2$ resulting in the creation of the three bosons with  the momenta $q_1^\prime$, $q_2^\prime$ and $q_3^\prime$. 
In addition, there is  a conjugate process that involves  two thermal bosons in the initial state that  produces the contribution of the same order of magnitude to the bosonic scattering  rate and modifies  the numerical prefactor in Eq.(\ref{Eq:sec2:tau_boson}). For the clarity of the presentation we limit our consideration here to the process of Fig.\ref{Fig:BosonicScattering}, restoring the correct prefactor at the end of the calculation.  

The decay rate corresponding to the process shown on  Fig. \ref{Fig:BosonicScattering} is given by
\begin{multline}
 \frac{1}{\tau_{q_1}(T)}=\frac{1}{3!}\int\frac{dq_2 dq_1^\prime dq_2^\prime dq_3^\prime}{(2\pi)^4}W_{q_1q_2}^{q_1^\prime, q_2^\prime, q_3^\prime}
 N_B(\omega_{q_2})\\\times\left[1+N_B(\omega_{q_1^\prime})\right]\left[1+N_B(\omega_{q_2^\prime})\right]\left[1+N_B(\omega_{q_3^\prime})\right].
\label{Eq:sec2:BosonTime1}
 \end{multline}
The integration  in Eq. (\ref{Eq:sec2:BosonTime1}) is limited to positive momenta; $N_B(\omega)$ stands for the Bose distribution function at temperature $T$ and 
\begin{multline}
 W_{q_1q_2}^{q_1^\prime q_2^\prime q_3^\prime}=(2\pi)^2\left|\langle 0| b_{q_1^\prime}b_{q_2^\prime}b_{q_3^\prime} \hat{T} b_{q_1}^+b_{q_2}^+|0\rangle\right|^2
 \delta(P_{\rm i}-P_{\rm f})\\\times\delta(E_{\rm i}-E_{\rm f})
 \label{Eq:sec2:W}
\end{multline}
is the transition probability. The delta functions in Eq. (\ref{Eq:sec2:W}) represent the energy and momentum conservation, and the bosonic creation and annihilation operators are related to the Fourier components of density via
\begin{equation}
b_q=\sqrt{\frac{2\pi}{L q}}\rho_q\,, \quad b_q^+=\sqrt{\frac{2\pi}{L q}}\rho_{-q}\,,\quad q>0.
\end{equation}

To evaluate the required matrix element of the $\hat{T}$-matrix, we resort to the perturbation theory with the formal small parameter $1/m$. Straightforward analysis shows that to the lowest order in $1/m$ the part of the $\hat{S}$-operator responsible for the scattering process under consideration is given by 
\begin{multline}
 \hat{S}= -\frac{16\pi^5 i}{L^4m^3 u_0^2}\sum_{\bf q}\tilde{\Gamma}^{5, \rm eff}_{\bf q} :\prod_{i=1}^5\rho_{q_i}:_B
  \delta\left(\sum_{i=1}^5 u_{q_i}q_i \right)
  \label{Eq:sec2:S}
\end{multline}
with
\begin{multline}
\tilde{\Gamma}^{(5, \rm eff)}_{\bf q}=\Gamma^{(5)}_{\bf q}- \frac{6 p_1 u_0\Gamma^{(3)}_{q_1, q_2, p_1}\Gamma^{(4)}_{q_3, q_4, q_5, -p_1}}{u_{p_1} p_1+u_{q_1}q_1+u_{q_2}q_2}\\+
 \frac{27 u_0^2p_1p_2\Gamma^{(3)}_{q_1, q_2, p_1}\Gamma^{(3)}_{q_3, -p_1, -p_2}\Gamma^{(3)}_{q_4, q_5, -p_2}}
 {4\left(u_{p_1}p_1+u_{q_1}q_1+u_{q_2}q_2\right) \left(u_{p_2}p_2+u_{q_4}q_4+u_{q_5}q_5\right)}.
\label{Eq:sec2:Gamma5Eff}
 \end{multline}
The momenta $p_1$ and $p_2$ in Eq. (\ref{Eq:sec2:Gamma5Eff}) are fixed by the momentum conservation in the vertices $\Gamma^{(3)}_{\bf q}$ and $\Gamma^{(4)}_{\bf q}$; we also assume the symmetrization of the left-hand side of Eq. (\ref{Eq:sec2:Gamma5Eff}) with respect to the five momenta $q_i$. 

To proceed further, we recall that we are interested in low momenta $q\ll 1/l$. Therefore, we can perform an expansion of the $S$-operator (\ref{Eq:sec2:S}) in powers of $ql$.
To illustrate the procedure, let us consider the second term in Eq. (\ref{Eq:sec2:Gamma5Eff}). 
Neglecting the dependence of the interaction vertices on the momentum and retaining only the cubic term in the bosonic dispersion relation  $qu_q=u_0(q-l^2q^3)$,  we find the corresponding contribution to $\tilde{\Gamma}^{(5, {\rm  eff} )}$
\begin{equation}
 \delta\tilde{\Gamma}^{(5, {\rm eff})}\propto \frac{\gamma_{(3, 0)}\gamma_{(4, 0)}}{l^2}\sum_{\sigma(q_1,\ldots q_5)}\frac{1}{q_1 q_2}.
\label{Eq:sec2:deltaGamma}
 \end{equation}
 Here we made explicit the symmetrization over the permutations of the momenta $q_1\,, \ldots, q_5$.
At this point one could think that the correction $\delta\tilde{\Gamma}^{(5, {\rm eff})}$, Eq. (\ref{Eq:sec2:deltaGamma}), is parametrically large [in parameter $1/(q l)^2$, where $q$ is the typical momentum]  compared to  the first term in Eq. (\ref{Eq:sec2:Gamma5Eff}), $\tilde{\Gamma}^{(5)}_{\bf q}\approx \gamma_{(5, 0)}$.  However, this is not not true.
The point is that, within our approximation (retaining only the linear and the cubic terms in the bosonic dispersion), the $\delta$-function responsible for the energy conservation in Eq. (\ref{Eq:sec2:S}) becomes $\delta\left(l^2\sum_{i=1}^5 q_i^3\right)$. 
It is now easy to see that the amplitude (\ref{Eq:sec2:deltaGamma}) identically vanishes on the mass shell\cite{Remark:deltaGamma}.  
To get a non-zero contribution to the $\hat{S}$-operator from the second term in Eq. (\ref{Eq:sec2:Gamma5Eff}), one needs to take into account higher-order corrections coming from the momentum dependence of the vortices 
$\Gamma_{\bf q}^{(3)}$ and $\Gamma_{\bf q}^{(4)}$ as well as from the fifth-order term  in the bosonic dispersion relation [both in the energy denominator in Eq. (\ref{Eq:sec2:Gamma5Eff}) and 
in the energy conservation condition]. This brings about an additional factor of the order of $q^2l^2$. As a result, the contribution of the second term in Eq.(\ref{Eq:sec2:Gamma5Eff}) to the $S$-matrix turns out to be of the same order as  that of the first one. A similar consideration applies to the last term in Eq. (\ref{Eq:sec2:Gamma5Eff}). 

A straightforward although lengthy calculation  with the account of Eqs. (\ref{Eq:sec1:uq})-(\ref{Eq:sec1:gamma5}) leads now to a desired expression for the $\hat{S}$-operator to the lowest order in $ql$:
\begin{equation}
\hat{S}=-\frac{16\pi^5 i}{L^4m^3 u_0^2}\Gamma_{(5, \rm eff)}\sum_{\bf q} :\prod_{i=1}^5\rho_{q_i}:_B
  \delta\left(u_0 l^2\sum_{i=1}^5 q_i^3 \right).
  \label{Eq:sec2:SFinal}
\end{equation}
The constant $\Gamma_{(5, \rm eff)}$ can be expressed in terms of the parameters of the Hamiltonian (\ref{Eq:sec1:H}). The corresponding expression is cumbersome and is presented in Appendix \ref{App:I}.

The transition probability corresponding to the $S$-matrix (\ref{Eq:sec2:SFinal}) is 
 \begin{multline}
 W_{q_1q_2}^{q_1^\prime q_2^\prime q_3^\prime}=\frac{8(5!)^2 \pi^5}{m^6 u_0^5 l^2}\Gamma_{(5, \rm eff)}^2 q_1 q_2 q_1^\prime q_2^\prime q_3^\prime\\\times
 \delta(q_1+q_2-q_1^\prime-q_2^\prime-q_3^\prime)\delta\left(q_1^3+q_2^3-q_1^{\prime3}-q_2^{\prime3}-q_3^{\prime3}\right).
 \label{Eq:sec2:WFinal}
\end{multline}
Let us now assume that the momentum $q_1\gg T/u_0$.
Considering the kinematics of the scattering process, we observe that, in view of energy and momentum conservation, the momenta of two  of the bosons in the final state 
(say, $q_2^\prime$ and $q_3^\prime$ ) are of the order of thermal momentum while the momentum of 
the third boson is parametrically close to $q_1$
\begin{equation}
q_1-q_1^\prime \sim T^3/u_0^3 q_1^2.
\end{equation}
Evaluating now the the bosonic lifetime according to Eq. (\ref{Eq:sec2:BosonTime1}) under the assumption $q_1\gg T/u_0$ one finds that it does not depend on the bosonic momentum and is given by 
\begin{equation}
\frac{1}{\tau_{q_1\gg T/u_0}(T)}=c_B \frac{\Gamma_{(5, \rm eff)}^2 T^5}{m^6 u_0^{10} l^2} 
\label{Eq:sec2:tau_boson}
\end{equation}
with the numerical constant $c_B\approx 6.5 \times 10^4$, see Appendix \ref{App:I}. 

Equation (\ref{Eq:sec2:tau_boson}) leads also to the estimate for the lifetime of a thermal boson in our system
\begin{equation}
 \frac{1}{\tau_B(T)}\sim \frac{T^5}{m^6 u_0^{10} l^2}\ ,
 \label{Eq:sec2:tau_boson_final}
\end{equation}
up to a constant of order unity. Equations (\ref{Eq:sec2:tau_boson}) and (\ref{Eq:sec2:tau_boson_final}) are 
central results of this section. Equation (\ref{Eq:sec2:tau_boson_final}) determines the characteristic time scale for the equilibration in a 
 chiral single-channel quantum Hall edge at relatively high temperatures, $m l^2 T\gg 1$.
The $T^5$-scaling in Eq. (\ref{Eq:sec2:tau_boson_final}) agrees with the that found for the intrabranch equilibration 
rate in a Wigner crystal in Ref.  \onlinecite{apostolov13}.

\section{Lifetime of fermionic excitations}
\label{sec:sec3}
\begin{figure}
\includegraphics[width=220pt]{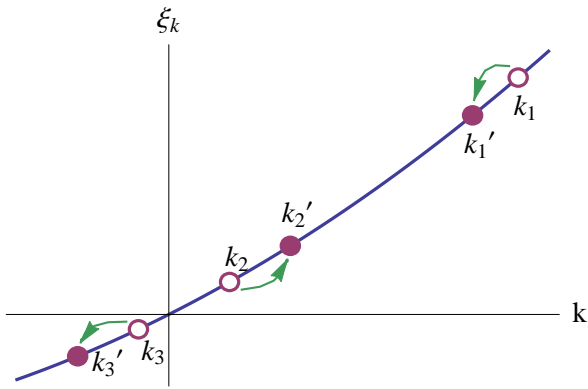}
\caption{\small The dominant fermonic scattering process contributing to the relaxation of a hot fermion with momentum $k_1\gg T/u_0$. The change in momentum of the hot fermion, 
$k_1-k_1^\prime \propto T^2/u_0 k_1$, is small compared to the typical momenta of the thermally excited fermions  $k_2$, $k_2^\prime$,  $k_3$ and $k_3^\prime$.  
 }
 \label{Fig:FermionicScattering}
\end{figure}

 We turn now to the investigation of fermionic excitations in our problem. 
 As we have discussed earlier, the fermions are expected to constitute the proper basis for the description of the  chiral edge at low temperatures, i.e.  $\lambda_T \equiv ml^2 T \ll 1$. 
Throughout this section we assume this condition to be satisfied. 
 
Our first task is to reexpress the Hamiltonian (\ref{Eq:sec1:H}) in the fermionic language. To achieve this goal, 
one plugs the fermionic representation (\ref{Eq:sec1:Refermionization}) of the the density components into Eq. (\ref{Eq:sec1:H}) and performs the normal ordering of the result with respect to fermions. 
The refermionization procedure results in the Hamiltonian of the form
\begin{equation}
 H=\sum_k \epsilon_k :a^+_ka_k:_F+\sum_{n=2}^{\infty} \tilde{H}^{(n)}.
 \label{Eq:sec3:HFermions}
\end{equation}
Here the first term corresponds to   free fermions with the dispersion relation $\epsilon_k$ while the other terms describe fermionic interactions
\begin{equation}
 \tilde{H}^{(n)}=\frac{1}{L^{n-1}}\sum_{\bf k, k^\prime}\Gamma^{(n)}_{\bf k, k^\prime}:\prod_{i=1}^n
 a_{k_{i}}^+a_{k^\prime_{i}}:_F.
\end{equation}
The fermionic vertices $\Gamma^{(n)}_{\bf k, k^\prime}$ are antisymmetric with respect to the momenta of incoming and outgoing particles,  ${\bf k}\equiv (k_1,\ldots k_n)$ and ${\bf k^\prime}\equiv 
(k_n^\prime, \ldots k_1^\prime)$. 

Let us discuss first the the structure of the fermionic Hamiltonian neglecting the bosonic vertices $\Gamma^{(n\geq 4)}_{\bf q}$ as well as the momentum dependence of the three-boson interaction $\Gamma^{(3)}_{\bf q}$. In this approximation all  fermionic vertices $\Gamma^{n\geq 3}_{\bf k, k^\prime}$ vanish and we are left with  fermions that have  dispersion relation
\begin{equation}
 \epsilon_k=\frac{3\gamma_{(3, 0)}}{2m}k^2+u_0\int_{0}^k dq \Gamma^{(2)}_q
\end{equation}
and  interact via two-body interaction
\begin{equation}
 \Gamma^{(2)}_{k_1, k_2, k^\prime_2, k_1^\prime}=\frac{\pi u_0}{2}\left(\Gamma^{(2)}_{k_1-k_1^\prime}-\Gamma^{(2)}_{k_1-k_2^\prime}\right).
 \label{Eq:sec3:Gamma2}
\end{equation}

We fix  $\gamma_{(3, 0)}=1/3$ by an appropriate choice of $m$ (see Ref.  \cite{Remark:Gamma3}).  The fermionic spectrum then reads
\begin{equation}
 \epsilon_k=u_0k+\frac{k^2}{2m}-\frac{u_0l^2k^3}{3}+O(u_0 l^4 k^5).
 \label{Eq:sec3:epsilon}
\end{equation}
We note that the third term in Eq. (\ref{Eq:sec3:epsilon}), representing  the renormalization of the fermionic spectrum  by the interaction\cite{Remark:ExchangeDiagram},  is small in the parameter $\lambda_T$. We postpone the discussion  of its effect till the end of this section.  

It is easy to see that, because of kinematic constraints, a two-fermion scattering process is not allowed \cite{Khodas2007}. To evaluate the lifetime of the fermionic excitations,  one thus needs to consider the three-particle collision process (see Fig.  \ref{Fig:FermionicScattering}) and the out-scattering part of the  corresponding  collision 
integral. The lifetime of a fermion at momentum $k_1\gg T/u_0$ is then given by
\begin{eqnarray}
 \frac{1}{\tau_{k_1}(T)} =\frac{1}{12} \int \frac{d\bf k}{\left(2\pi\right)^5}W_{k_1, k_2,
k_3}^{k_1^\prime, k_2^\prime, k_3^\prime}N_F(\epsilon_{k_2})
 N_F(\epsilon_{k_2}) \nonumber \\
 \times \left(1-N_F(\epsilon_{k_1^\prime})\right)\left(1-N_F(\epsilon_{k_2^\prime})\right)  \left(1-N_F(\epsilon_{k_3^\prime})\right).
 \label{Eq:sec3:tau_fermion}
\end{eqnarray}
Here the transition probability $W_{k_1, k_2, k_3}^{k_1^\prime, k_2^\prime, k_3^\prime}$ can be expressed in terms of the matrix element for the three-particle collision process
\begin{eqnarray}
 W_{k_1, k_2, k_3}^{k_1^\prime, k_2^\prime, k_3^\prime} &=& (2\pi)^2
\left|\langle 1, 2, 3|T| 1^\prime, 2^\prime,
3^\prime\rangle\right|^2 \nonumber \\
&\times& \delta\left(E_{\mathrm i}-E_{\mathrm
f}\right) \delta\left(P_{\mathrm i}-P_{\mathrm f}\right)
 \,,
 \label{Eq:sec3:W}
\end{eqnarray}
with $\delta$-functions ensuring the momentum and energy conservation. 

To evaluate the matrix element for the three-particle scattering, we employ the lowest order perturbation theory in the fermionic interaction (\ref{Eq:sec3:Gamma2}), yielding
\begin{eqnarray}
 \langle 1, 2, 3|T| 1^\prime, 2^\prime,
3^\prime\rangle=\frac{4 \cdot (3!)^2\Gamma^{(2)}_{k_1, k_2, k2^\prime, q}\Gamma^{(2)}_{q, k_3, k_3^\prime, k_1^\prime}}{\epsilon_{k_1}+\epsilon_{k_2}-\epsilon_{p_2^\prime}-\epsilon_q}.
\label{Eq:sec3:T}
\end{eqnarray}
Here we assume the antisymmetrization of the right-hand side with respect to the incoming and outgoing momenta. 

To proceed further, we expand the fermionic interaction $\Gamma^{(2)}_{\bf k, \bf k^\prime}$ in powers of momentum,
 see Eqs. (\ref{Eq:sec1:Gamma2}), (\ref{Eq:sec3:Gamma2}). 
The  antisymmetrization in Eq.(\ref{Eq:sec3:T}), required by the Fermi statics of our particles,  has a profound impact on the result. Indeed, assuming that the matrix element remains analytic at small momenta on the  mass shell, one immediately concludes that $\langle 1, 2, 3|T| 1^\prime, 2^\prime,
3^\prime\rangle$ is proportional to the sixth power of momentum
\begin{equation}
 \langle 1, 2, 3|T| 1^\prime, 2^\prime,3^\prime\rangle\propto\prod_{i>j}(k_i-k_j)(k_i^\prime-k_j^\prime) .
\end{equation}
In particular, to obtain a non-zero result from Eq. (\ref{Eq:sec3:T}) one needs to expand the product of two-particle interactions in the numerator to the eighth power of momentum. 
The contributions of all the lower expansion terms vanish\cite{Remark:power8}. As a result we find
\begin{eqnarray}
 \langle 1, 2, 3|T| 1^\prime, 2^\prime,3^\prime\rangle&=&12\pi^2u_0^2 m l^8 \left(10 \gamma_{(2, 4)}^2+7\gamma_{(2, 6)}\right)\nonumber\\
 &\times&\prod_{i>j}(k_i-k_j)(k_i^\prime-k_j^\prime)
\label{Eq:sec3:TFinal}
\end{eqnarray}

It is now easy to calculate the fermionic lifetime, Eq.(\ref{Eq:sec3:tau_fermion}). We shall take into account that for $k_1\gg T/u_0$ the kinematics of the process requires that
one of the outgoing momenta (say $k_1^\prime$) is close to $k_1$,
\begin{equation}
 k_1-k_1^\prime\propto T^2/k_1.
\end{equation}
Evaluating the integral we get
\begin{equation}
 \frac{1}{\tau_{k_1\gg T/u_0}(T)}=\frac{c_F \left(10 \gamma_{(2, 4)}^2+7\gamma_{(2, 6)}\right)^2k_1^7 T^7 m^3 l^{16}}{u_0^{10}},
 \label{tau_k_fermion}
\end{equation}
with a numerical constant $c_F\approx 2.4\times 10^4$ (see Appendix \ref{App:I}). 

Setting $k_1\sim T/u_0$, we obtain an estimate (up to a numerical coefficient of order unity) for the life time of a fermionic excitation with an energy of the order of temperature:
\begin{equation}
 \frac{1}{\tau_F(T)}\sim \frac{T^{14} m^3 l^{16}}{u_0^{10}}.
\label{Eq:sec3:tauF}
\end{equation}
Equations  (\ref{tau_k_fermion}) and (\ref{Eq:sec3:tauF}) constitute the central results of this section. Equation (\ref{Eq:sec3:tauF}) determines the equilibration rate 
of chiral single-edge quantum Hall edge channel at low temperatures, $m l^2 T\ll 1$. 

In our discussion we have so far neglected higher-order scattering processes, the higher-order bosonic vertices $\Gamma^{(n)}_{\bf q}$, as well as the exchange corrections to the fermionic single-particle spectrum [the third term in Eq. (\ref{Eq:sec3:epsilon})]. Let us now discuss the stability of the result (\ref{Eq:sec3:tauF})  with respect to these corrections. First we note, that the $T^{14}$ scaling   of the fermionic  relaxation rate is extremely robust, as it relies solely on the 
$k^6$ scaling of the three-particle collision amplitude dictated by the Pauli principle. Estimating contributions of higher-order scattering processes (four-particle etc.), one obtains still higher powers of $T$, with $\lambda_T = ml^2T$ being a dimensionless small parameter of the expansion. Further, 
the contributions of many-boson interaction vertices to the three-particle collision rate are suppressed by positive powers of the
parameter $1/m u_0 l\lesssim1$. To  illustrate this point, let consider the three-boson interaction vertex $\Gamma^{(3)}_{\bf q}$. It contributes to the three-fermion collision amplitude already in the first order perturbation theory. The corresponding contribution reads
\begin{equation}
 \langle 1, 2, 3|T| 1^\prime, 2^\prime,3^\prime\rangle_{\Gamma^{(3)}_{\bf q}}\propto 
\frac{l^6}{m}\prod_{i>j}(k_i-k_j)(k_i^\prime-k_j^\prime)
\label{three-boson-correction}
\end{equation}
and thus contains an additional factor $1/(m u_0 l)^2$ in comparison with Eq. (\ref{Eq:sec3:TFinal}). If $m u_0 l \sim 1$ (in which case only the fermonic regime is realised), 
Eq.~(\ref{three-boson-correction})  is of the same order as Eq. (\ref{Eq:sec3:TFinal}), thus modifying the corresponding numerical prefactor. On the other hand,  if $m u_0 l \gg 1$ (which is the condition for the existence of the high-temperature bosonic regime), the contribution of the three-boson interaction vertex, Eq.~(\ref{three-boson-correction}), yield only  a small correction to Eq. (\ref{Eq:sec3:TFinal}).

 The situation with the exchange correction to the fermionic spectrum is slightly more subtle. One can try to incorporate it  into the denominator in Eq. (\ref{Eq:sec3:T}) 
and the $\delta$-function expressing the energy conservation in Eq. (\ref{Eq:sec3:W}).
Performing then the expansion of the thee-particle collision amplitude at the mass shell in powers of momentum, one finds
\begin{equation}
 \langle 1, 2, 3|T| 1^\prime, 2^\prime,3^\prime\rangle_\Sigma\propto 
l^{10} u_0^4 m^3\prod_{i>j}(k_i-k_j)(k_i^\prime-k_j^\prime).
\label{Eq:sec3:TSigma}
\end{equation}
We see that the $k^6$ scaling of the collision amplitude (responsible for the $T^{14}$ scaling of the relaxation rate) is preserved, as expected. On the other hand, the contribution (\ref{Eq:sec3:TSigma}) appears to be large compared to Eq. (\ref{Eq:sec3:TFinal}) 
in the parameter $(u_0ml)^2$. The two additional powers of mass $m$ in Eq. (\ref{Eq:sec3:TSigma}) come from the second order expansion of the  amplitude (\ref{Eq:sec3:T}) in $\delta\epsilon_k=-l^2k^3/3$. 
However, in accordance with the rules of a diagrammatic expansion, the self-energy correction produced by  the third term in Eq. (\ref{Eq:sec3:epsilon}) cannot be considered separately from the corresponding  vertex corrections it generates to Eq. (\ref{Eq:sec3:T}). We expect that the correct account of  the vertex corrections (which is a very tedious task for the three-particle processes under consideration) would lead to the cancellation of (\ref{Eq:sec3:TSigma}), so that Eq. (\ref{Eq:sec3:TFinal}) yields the leading contribution to the three-particle scattering amplitude, including the prefactor. 

\section{Summary and discussion}
\label{sec4}

In this work, we have presented a detailed analysis of the relaxation process in a chiral single-channel quantum Hall edge state (integer or fractional). The relaxation is naturally described in the fermionic language at low temperatures, $\lambda_T \ll 1$,  and in the bosonic language at high temperatures, $\lambda_T \gg 1$, where $\lambda_T=ml^2T$ is the dimensionless parameter of the theory. The relaxation rates of a hot excitation with momentum $k$ scale as $T^5k^0$, Eq. (\ref{Eq:sec2:tau_boson}) for a boson and as $T^7k^7$ for a fermion (where $k$ is counted from Fermi momentum), Eq.~(\ref{tau_k_fermion}). The equilibration rate of the system behaves as $T^5$ in the high-temperature (bosonic) regime and as $T^{14}$ in the low-temperature (fermionic) regime, as expressed by Eqs. (\ref{Eq:sec2:tau_boson_final}) and (\ref{Eq:sec3:tauF}). These results match at $\lambda_T\sim 1$ where a Bose-Fermi crossover takes place. 

Before closing the paper, we discuss the connection of our findings with related recent advances in the field:

\begin{enumerate}
\item[i)]
It is instructive to compare the obtained relaxation rates with those in an analogous but non-chiral system, Ref.~\onlinecite{Protopopov2014_2}. When both left- and right-moving excitations branches are present, the relaxation rates are  $\sim T^5/m^4 u_0^8$ and  $\sim l^4 T^7 / m^2 u_0^8$ in the bosonic and fermionic regimes, respectively.  Thus, in the bosonic regime the temperature scaling is $T^5$ in both chiral and non-chiral cases; the only difference is in the additional factor $(mu_0l)^{-2}\lesssim 1$ in the chiral case. In the fermionic regime the chirality of the system has a much more dramatic impact: it changes the temperature scaling of the relaxation rate from $T^7$ to $T^{14}$. Such a strong suppression of the relaxation can be traced back to a very strong effect of antisymmetrization for a three-body fermion scattering in a single-channel chiral system.

\item[ii)]
It is worth reminding that we have assumed a finite-range interaction, with a sufficient degree of analyticity at small momenta.
A comparison with the results obtained in a perturbative fermionic calculation for a chiral Luttinger liquid with unscreened ($1/r$) Coulomb interaction indicates  \cite{Karzig2012} that  the finite-range character of the interaction strongly suppresses the relaxation in the fermionic regime. The Bose-Fermi ''phase diagrams'' for a chiral quantum Hall edge with long-range interactions remains to be explored.

\item[iii)]  It was shown in Refs. \onlinecite{Protopopov13, Protopopov14} that solitonic
density waves  arise in the course of time evolution of a
strong enough density pulse (on top of zero-temperature
background) in the bosonic regime $\lambda_E \gg 1$, where $E$ is set by the pulse amplitude.
It was argued in a recent preprint \cite{PustilnikMatveev2014} that, upon quantisation, solitons of the bosonic theory can be regarded as  a continuation of the particle branch of fermionic excitations to the bosonic part of the phase space
(while the standard bosonic branch constitutes the continuation of the
hole fermionic branch). The authors of Ref.~\onlinecite{PustilnikMatveev2014} focus on a non-chiral system characterised by the Bose-Fermi duality established in Ref. \onlinecite{Protopopov2014_2}. One can expect, however, that this conjecture would be equally applicable to a chiral system. It would be very interesting to verify these conjectures and to calculate the life time of such soliton-like excitations in the bosonic regime, both in the chiral and non-chiral cases. It should be emphasised, however, that the corresponding life times are expected to be much shorter than those of bosons. Therefore, they will not influence the long-time equilibration rates discussed in Ref. \onlinecite{Protopopov2014_2} (for the non-chiral case) and in the present work (for a chiral system). 

\item[iv)] The energy relaxation in quantum Hall edges (both chiral and non-chiral) was probed in the recent experiments, Refs. \onlinecite{Altimiras10, Paradiso11, Deviatov11, Prokudina2014}. We expect that further development of the methods employed in those works  will allow systematic investigation of the temperature (or energy) dependence of the equilibration rates.
\end{enumerate}

\begin{acknowledgments}

We acknowledge useful discussions with A. Levchenko and financial support by DFG Priority Program 1666, by Israeli Science Foundation,  by German-Israeli Foundation, and by the EU Network Grant InterNoM. The research of A.D.M. was supported by the Russian Science Foundation (project 14-22-00281).

\end{acknowledgments}

\vskip0.5cm
 
\appendix
\section{Expressions for numerical prefactors}
\label{App:I}

\begin{widetext}
In this Appendix we collect results for numerical prefactors in the formulas for the lifetime of the bosonic  and fermionic excitations. (The corresponding expressions are quite lengthy and for this reason have not been included in the main text.)

The constant $\Gamma_{(5, \rm eff)}$   in Eqs.~(\ref{Eq:sec2:SFinal}),  (\ref{Eq:sec2:SFinal}), and in the final result for the hot-boson relaxation rate (\ref{Eq:sec2:tau_boson}) has the following expression in terms of the parameters of the Hamiltonian (\ref{Eq:sec1:H}):
\begin{multline}
  \Gamma_{(5, \rm eff)}= \gamma_{(5, 0)}
  +2\left(\gamma_{(2, 4)}\gamma_{(3, 0)}\gamma_{(4, 0)}+\gamma_{(3, 2)}\gamma_{(4, 0)}+2\gamma_{(3, 0)}\gamma_{(4, 2)}\right)\nonumber\\+
  \frac{\gamma_{(3, 0)}}{6}\left(20\gamma_{(2, 4)}^2\gamma_{(3, 0)}^2+14\gamma_{(2, 6)}\gamma_{(3, 0)}^2+67\gamma_{(2, 4)}\gamma_{(3, 0)}\gamma_{(3, 2)}
  +48\gamma_{(3, 2)}^2+66\gamma_{(3, 0)}\gamma_{(3, 4)}\right)
\end{multline}

The numerical constant $c_B$ in the expression for the bosonic lifetime,  Eq. (\ref{Eq:sec2:tau_boson}), is given by the following dimensionless integral
\begin{equation}
c_B=\frac{(5!)^2\pi}{12}\int d\omega_1 d\omega_2  \omega_1\omega_2(\omega_1+\omega_2)
\left[\tilde{N}_B(\omega_1+\omega_2)[1+\tilde{N}_{B}(\omega_1)]
[\tilde{N}_B(\omega_2)+1]+\tilde{N}_B(\omega_1)\tilde{N}_B(\omega_2)[1+\tilde{N}_B(\omega_1+\omega_2)]\right],
\label{Eq:app:c}
\end{equation}
where $\tilde{N}_B(\omega)$ is a dimensionless Bose distribution with the temperature $T=1$ and the chemical potential $\mu=0$.
The first term in the square brackets represents the contribution of the scattering process shown on Fig. \ref{Fig:BosonicScattering}, while the second term corresponds to the conjugate process with two thermal bosons in the initial state. 

The numerical constant $c_F$ in the expression for the fermionic lifetime,  Eq. (\ref{tau_k_fermion}), is given by the following dimensionless integral
\begin{equation}
 c_F=\frac{9}{2\pi}\int d\omega_1 d\omega_2 d\omega_1^\prime d\omega_2^\prime (\omega_1-\omega_2)^2(\omega_1^\prime-\omega_2^\prime)^2\tilde{N}_F(\omega_1)\tilde{N}_F(\omega_2)\left[1-\tilde{N}_F(\omega_1^\prime)\right]\left[1-\tilde{N}_F(\omega_2^\prime)\right]\delta(\omega_1+\omega_2-\omega_1^\prime-\omega_2^\prime),
\end{equation}
where $\tilde{N}_F(\omega)$ is a dimensionless Fermi distribution with the temperature $T=1$ and the chemical potential $\mu=0$.

\end{widetext}


\begin{thebibliography}{100}

\bibitem{Mason09} Y.-Fu Chen, T. Dirks, G. Al-Zoubi, N. O. Birge, and N. Mason, Phys. Rev. Lett. {\bf 102}, 036804 (2009).

\bibitem{Yacoby10} G. Barak, H. Steinberg, L.N. Pfeiffer, K.W. West, L.  Glazman, F. von Oppen, and A. Yacoby, Nature Phys. {\bf 6}, 489 (2010). 

\bibitem{Altimiras10}  
C. Altimiras, H. le Sueur, U. Gennser, A. Cavanna, D. Mailly, and F. Pierre, Nature Phys. {\bf 6}, 34 (2010); Phys. Rev. Lett. {\bf 105}, 226804 (2010);
H. le Sueur, C. Altimiras, U. Gennser, A. Cavanna, D. Mailly, and F. Pierre, Phys. Rev. Lett. {\bf 105}, 056803 (2010).

\bibitem{Paradiso11} 
N. Paradiso, S. Heun, S. Roddaro, L. Sorba, F. Beltram, and G. Biasiol, Phys. Rev. B {\bf 84}, 235318 (2011).

\bibitem{Deviatov11} 
E.V. Deviatov, A. Lorke, G. Biasiol, and L. Sorba, Phys. Rev. Lett. {\bf 106}, 256802 (2011).

\bibitem{Prokudina2014} M.G. Prokudina, S. Ludwig, V. Pellegrini, L. Sorba, G. Biasiol, and V.S. Khrapai, Phys. Rev. Lett.  {\bf 112}, 216402 (2014). 

\bibitem{Kinoshita06} 
T. Kinoshita, T. Wenger, and D.S. Weiss, Nature {\bf 440}, 900 (2006).

\bibitem{Hofferberth07} 
S. Hofferberth, I. Lesanovsky, B. Fischer, T. Schumm, and J. Schmiedmayer,  Nature {\bf 449}, 324 (2007).

\bibitem{Stone_book} M. Stone, \textit{Bosonization} (World Scientific, 1994).

\bibitem{Delft} J. von Delft and  H. Schoeller,
Annalen Phys. {\bf 7}, 225  (1998).

\bibitem{Gogolin}
A.O. Gogolin, A.A. Nersesyan, and A.M. Tsvelik,
\textit{Bo\-son\-iza\-tion in Strongly Correlated Systems},
(University Press, Cambridge 1998).

\bibitem{Giamarchi}
T. Giamarchi, \textit{Quantum Physics in One Dimension},  (Claverdon
Press Oxford, 2004).

\bibitem{Matis_Lieb1965}
D.C. Mattis and E.H. Lieb, J. Math. Phys. {\bf 6}, 304 (1965).

\bibitem{Rozhkov}
A.V. Rozhkov, Phys. Rev. B {\bf 77}, 125109 (2008); Phys. Rev. B {\bf 74},
245123 (2006); Eur. Phys. J. {\bf 47} , 193 (2005).

\bibitem{imambekov09} A.~Imambekov and L.I.~Glazman, Science {\bf 323}, 228
(2009); Phys. Rev. Lett. {\bf 102}, 126405 (2009).

\bibitem{imambekov11} A.~Imambekov, T.L.~Schmidt, and L.I.~Glazman, Rev. Mod.
Phys {\bf 84}, 1253 (2012).

\bibitem{Lunde07} A.M.~Lunde, K.~Flensberg, and L.I. Glazman, Phys. Rev. B
{\bf 75}, 245418 (2007).

\bibitem{Khodas2007} M. Khodas, M. Pustilnik, A. Kamenev, L.I. Glazman,  Phys.
Rev. B {\bf 76}, 155402 (2007).

\bibitem{micklitz11} T.~Micklitz and A.~Levchenko, Phys. Rev. Lett. {\bf 106},
196402 (2011). 

\bibitem{ristivojevic13} Z.~Ristivojevic and K.A.~Matveev, Phys. Rev. B {\bf
87}, 165108 (2013). 

\bibitem{Dmitriev2012}
A.P. Dmitriev, I.V. Gornyi, and D.G. Polyakov, Phys. Rev. B {\bf 86}, 245402 ( 2012).



\bibitem{MatveevFurusaki2013} K.A. Matveev, A. Furusaki,  Phys. Rev. Lett. {\bf
111}, 256401 (2013).

\bibitem{Gangardt2014} M. Arzamasovs, F. Bovo, and D.M. Gangardt, Phys. Rev. Lett. {\bf  112}, 170602 (2014)

\bibitem{Protopopov14} I.V. Protopopov, D.B. Gutman, M. Oldenburg, and A.D. Mirlin, Phys. Rev. B {\bf 89}, 161104(R) (2014).

\bibitem{apostolov13} S.~Apostolov, D.E.~Liu, Z.~Maizelis, and
A.~Levchenko, Phys. Rev. B {\bf 88}, 045435 (2013). 

\bibitem{Lin2013} J. Lin, K. A. Matveev, M. Pustilnik,  Phys. Rev. Lett. {\bf
110}, 016401 (2013).

\bibitem{Protopopov2014_2}
I.V. Protopopov, D.B. Gutman, A.D. Mirlin,  Phys. Rev. B {\bf 90}, 125113 (2014) 

\bibitem{schick68} M.~Schick, Phys. Rev. {\bf 166}, 404 (1968). 


\bibitem{Protopopov13} I.V. Protopopov, D.B. Gutman, P. Schmitteckert, and A.D. Mirlin, Phys. Rev. B {\bf 87}, 045112, (2013).

\bibitem{heiblum}
Y. Ji, Y.C. Chung, D. Sprinzak, M. Heiblum, D. Mahalu, and
H. Shtrikman, Nature (London) {\bf 422}, 415 (2003);
I. Neder, M. Heiblum, Y. Levinson, D. Mahalu, and V. Umansky,
Phys. Rev. Lett. {\bf 96}, 016804 (2006);
I. Neder, M. Heiblum, D. Mahalu, and V. Umansky,
Phys. Rev. Lett. {\bf 98}, 036803 (2007);
I. Neder, F. Marquardt, M. Heiblum, D. Mahalu, and V. Umansky,
Nat. Phys. {\bf 3}, 534 (2007).



\bibitem{schoenenberger}
E. Bieri, M. Weiss, O. G\"oktas, M. Hauser, C. Sch\"onenberger, and
S. Oberholzer, Phys. Rev. B {\bf 79}, 245324 (2009). 

\bibitem{roche}
P. Roulleau, F. Portier, D.C. Glattli, P. Roche, A. Cavanna,
G. Faini, U. Gennser, and D. Mailly, Phys. Rev. B {\bf 76}, 161309(R)
(2007);
P. Roulleau, F. Portier, D.C. Glattli, P. Roche, A. Cavanna,
G. Faini, U. Gennser, and D. Mailly, Phys. Rev. Lett. {\bf 100}, 126802
(2008);
P. Roulleau, F. Portier, P. Roche, A. Cavanna, G. Faini, U. Gennser,
and D. Mailly,  Phys. Rev. Lett. {\bf 101}, 186803 (2008);
{\it ibid} {\bf 102}, 236802 (2009);
P.-A.~Huynh, F.~Portier, H.~le~Sueur, G.~Faini, U.~Gennser, D.~Mailly, F.~Pierre, W.~Wegscheider, and P.~Roche, Phys. Rev. Lett. {\bf 108}, 256802 (2012).
 
\bibitem{strunk}
L.V. Litvin, H.-P. Tranitz, W. Wegscheider, and C. Strunk,
Phys. Rev. B {\bf 75}, 033315 (2007);
L.V. Litvin, A. Helzel, H.-P. Tranitz, W. Wegscheider, and
C. Strunk, Phys. Rev. B {\bf 78}, 075303 (2008); {\it ibid} {\bf 81},
205425 (2010).


\bibitem{Wen} X.G. We,  Phys. Rev. B {\bf 41}, 12838 (1990); Phys. Rev. Lett. {\bf 64}, 2206 (1990);  Phys. Rev. {\bf B 43},
11025 (1991)

 \bibitem{Remark:CommutatioRelation} For the Laughlin $\nu=1/m$ states our normalization of chiral boson $\rho(x)$ is somewhat non-standard. 
In our notations the fermionic and quasiparticle creation operators are give by $\exp\left(i\sqrt{1/\nu}\phi\right)$ and $\exp\left(i\sqrt{\nu}\phi\right)$ respectively. 
 The physical charge denisty on the edge   $\tilde{\rho}(x)=\sqrt{\nu}\rho(x)=\sqrt{\nu}\partial_x\phi/2\pi$. 
 
















 \bibitem{Remark:BoseFermiMapping} In the case of the integer $\nu=1$ quantum Hall effect, the operators $a_k^+$ and $a_k$ essentially create and annihilate the physical fermions comprising the 
 quantum Hall state. On the other hand, for the  $\nu=1/m$ Laughlin states, fermions $a^+_k$ and $a_k$ create and annihilate fermionic quasiparticles (''composite fermions'') that are very much different from the original electrons. 
 
 \bibitem{Remark:Gamma3} The value of the of the vertex $\Gamma^{(3)}_{\bf q}$ at zero mementum can be fixed to  $\Gamma^{(3)}_{\bf q=0}=1/3$ by the sutable redeffinition of the parameter $m$ in Eq.(\ref{Eq:sec1:Hn}). We note also that arbitrary uniform fourth-power  symmetric polynomial of the theree variables $(q_1, q_2, q_3)$ can be expessed as a linear combination of the 
 four independent symmetric polynomials $p_{3, 4}^{(1)}=\left(\sum_{i=1}^3q_i\right)^4$, $p_{3, 4}^{(2)}=\left(\sum_{i=1}^3q_i\right)^2\left(\sum_{j=1}^3q_j^2\right)$, 
 $p_{3, 4}^{(3)}=\left(\sum_{i=1}^3q_i\right)\left(\sum_{j=1}^3q_j^3\right)$ and $p_{3, 4}^{(4)}=\left(\sum_{i=1}^3q_i^2\right)^2$. The contribution of the first $3$ of these  polynomilas
 to $\Gamma^{(3)}_{\bf q}$ vanishes due to momentum conservation condition $\sum_{i=1}^3 q_i=0$. 
 
 \bibitem{Remark:deltaGamma} Indeed, the sum over permutattions in Eq. (\ref{Eq:sec2:deltaGamma}) is proportional to $(s_1^3-3s_1 s_2+2 s_3)/q_1 q_2 q_3 q_4 q_5$ where $s_n$ are symmetric polynomials $s_n=\sum_{i=1}^5 q_i^n$.
 
\bibitem{Remark:ExchangeDiagram} In the language of diagrammatics the cubic correction to the fermionic spectrum as well as the other terms denoted by $O(u_0 l^4 k^5)$ 
in Eq. (\ref{Eq:sec3:epsilon}) corespond to the first-order  exchange diagram of the zero-temperature perturbation theory.   

\bibitem{Remark:power8} We have also checked this by direct calculation. 

 

\bibitem{PustilnikMatveev2014}
M. Pustilnik and K.A. Matveev, arXiv:1409.6721.


\bibitem{Karzig2012} T. Karzig, A. Levchenko, L.I. Glazman, and  F. von Oppen,  New J. Phys. {\bf 14}, 105009 (2012).




\end{thebibliography}
\end{document}